\newif\ifAMStwofonts
\AMStwofontstrue

\def\gsim{~\rlap{$>$}{\lower 1.0ex\hbox{$\sim$}}}

\def\simpropto{\lower.2ex\hbox{$\; \buildrel \propto \over \sim \;$}}
\def\ltsim{\lower.5ex\hbox{$\; \buildrel < \over \sim \;$}}
\def\gtsim{\lower.5ex\hbox{$\; \buildrel > \over \sim \;$}}
\def\ltsim{\lower.5ex\hbox{$\; \buildrel < \over \sim \;$}}
\def\gtsim{\lower.5ex\hbox{$\; \buildrel > \over \sim \;$}}

\newcommand{\beq}{\begin{equation}}
\newcommand{\eeq}{\end{equation}}
\def\fixit#1{}

\documentclass[12pt,aams4,notitlepage]{emulateapj}
\usepackage{natbib}
\usepackage{color}
\def\beq{\begin{equation}}
\def\eeq{\end{equation}}
\def\bea{\begin{eqnarray}}
\def\eea{\end{eqnarray}}
\def\simlt{\lower.5ex\hbox{$\; \buildrel < \over \sim \;$}}
\def\simgt{\lower.5ex\hbox{$\; \buildrel > \over \sim \;$}}
\def\simpropto{\lower.2ex\hbox{$\; \buildrel \propto \over \sim \;$}}

\begin{document}

\title
{Unleashing Positive Feedback: Linking  the Rates of Star Formation, Supermassive Black Hole Accretion and Outflows
in Distant Galaxies}

\author{ Joseph Silk}
\affil{
Institut d'Astrophysique,  UMR 7095 CNRS, Universit\'e Pierre et Marie Curie, 98bis Blvd Arago, 75014 Paris, France\\
Department of Physics and Astronomy, The Johns Hopkins University,
Homewood Campus, Baltimore MD 21218, USA\\
Beecroft Institute of Particle Astrophysics and Cosmology, Department of Physics, University of Oxford, Oxford OX1 3RH, UK
}

\begin{abstract}
Pressure-regulated star formation is a simple variant on the usual supernova-regulated star formation efficiency that controls the global star formation rate as a function of cold gas content in star-forming galaxies, and accounts for the Schmidt-Kennicutt law in both nearby and distant galaxies. Inclusion of AGN-induced pressure, by jets and/or winds that flow back onto  a gas-rich disk, can lead   under some circumstances to significantly enhanced star formation rates, 
especially at high redshift and most likely followed by the more widely accepted phase of star formation quenching. Simple expressions are derived that relate supermassive black hole growth, star formation and outflow rates. The ratios of black hole to spheroid mass and  of both black hole accretion and outflow rates to star formation rate are predicted as a function of time. I suggest various tests of the AGN-triggered star formation hypothesis.
\end{abstract}
\maketitle

\section{Introduction}
One of the great mysteries in galaxy formation is the connection with supermassive black holes and AGN. 
AGN outflows are generally thought to play an important role in quenching star formation, as in \cite{nesvadba10} and references cited therein. 
 An impressive recent example of negative feedback driven by a quasar on galactic scales is given by \cite{maiolino12}.

In rare, usually nearby, cases, there is evidence for  triggering of star formation. How prevalent this might be at early epochs is unknown. AGN luminosities and nuclear star formation rates  are  correlated \citep{imanishi11}. Jet-triggering of star formation in nearby galaxies \citep{croft06, zaurin07, crockett12}, of distant galaxies \citep {feain07, elbaz09}
and even of molecular hydrogen cloud formation at high redshift \citep{klamer04},
is observed.

There is a quantitative connection between black hole  mass and spheroid velocity dispersion ($M_{BH} \simpropto \sigma^{4-5}$) with approximate bulge mass  scaling $M_{BH}\approx 10^{-3}M_{spheroid}$  (\cite{graham11} and references cited therein), with deviations at low \citep{graham12} and high masses \citep{mcconnell12}. There is also a relation  between black hole accretion rate and star formation rate ($\dot M_{BH}\approx 2.10^{-3}\dot M_{\ast}$) in AGN \citep{silverman09, mullaney12}. Rapid black hole growth and intense host galaxy star formation are inferred to be coeval at $z\sim 6$  \citep{wang11}.  AGN-heated molecular disks  are responsible for black hole growth and show little evidence for star formation \citep{sani12}, which however dominates further out. For example, an interesting example of coexisting black hole growth and star formation is the case of  IRAS 20551- 4250 \citep{sani11},
where the AGN-heated molecular disk dominates the emissivity at mid-IR wavelengths on $\sim 100$ pc scales but star formation dominates at NIR and FIR wavelengths on kpc scales.

  {  
In this paper, I revisit the theory of  AGN-triggered star formation, and develop an analytical formulation intended to complement recent simulations, most notably those of \cite{gaibler12}. The theory of AGN-triggered star formation was developed  in \cite{silk09}.
Subsequent applications include 1-d \citep{ishibashi12}, 2-d \citep{liugan13} and  3-d numerical simulations  in homogeneous halos \citep{zubovas13}. However  only the 3-d  studies  of  propagation of  jets   \citep{gaibler12, wagner12} and winds \citep{wagner13} in inhomogeneous halos   capture the appropriate physics of outflows  and induced star formation.
}
 
   {  
  One motivation for introducing AGN outflows (jets or winds) as a star formation trigger is that a time-scale naturally arises that is shorter than the gravitational time-scale \citep{silk09}. I argue below that a shorter time-scale at high redshift is motivated observationally. Another is that radiation-pressure driven outflows are  generically leaky because of Rayleigh-Taylor instabilities and fail  to provide the momentum needed via multiple scatterings to account for the $M_{BH}-\sigma$ scaling relation \citep{silknusser10, krumholz13}.
  }
 
 The theory  is intended to provide a second mode of star formation, dominant at high redshift and rare at low redshift, that can account for the observed elevated, and more slowly varying at high redshift, specific star formation rate (SSFR), defined by $SSFR=\dot M_\ast/M_\ast$  (see \cite{weinmann11} for a recent compilation of the data, and \cite{bouwens12, stark13} for updates with revised dust and nebular emission corrections, respectively) in luminous star-forming galaxies. 
The star formation rate phenomenology is outlined in Section 2 and the mechanism for positive feedback  is developed in Section 3, where
I account for the interconnection between black hole growth, star formation and outflow rates.  A final section provides a summary and discussion of possible tests of the AGN positive feedback hypothesis.

\section {Star formation rate}
My starting point is the well-tested phenomenological expression for 
the star formation rate in disk galaxies, 
motivated by  the instability of self-gravitating cold gas-rich disks to non-axisymmetric instabilities that lead to  fragmentation and  giant molecular cloud formation, with subsequent  star formation modulated by feedback from supernovae. This simple formalism  fits star-forming  galaxy data on the Kennicutt-Schmidt law, both nearby and at $z\sim 2$ \citep{kennicutt98,Genzel+10},
and even  fits data on individual star-forming cloud complexes provided the local free-fall time is used instead of $1/\sqrt {G\rho_d}$ \citep{krumholz12}, where $\rho_d$ is the mean disk  density. 
  {   I note in passing that not all authors agree on the role of supernova feedback as the primary culprit that controls the normalisation of the K-S relation.
However supernova feedback as an explanation of galactic star formation in efficiency is certainly a majority viewpoint.}

One may write the star formation rate per unit area in a galactic disk as 
 $$
\dot\Sigma_\ast=\epsilon_{SN}
\Sigma_{gas}  /t_{dyn},
$$
where $\Sigma_{gas} $ is the gas surface density at a disk scale length $R_d$ and $t_{dyn}$ is a disk dynamical time, taken here to be $R_d/\sigma, $ with $\sigma$ taken to be the circular velocity.
I define the gas pressure in the disk, appropriate for interstellar  clouds of velocity dispersion $\sigma_d,$  by 
{{}
$p_{d,gas}=\rho_{d,gas} \sigma_d^2, $
}
 where $\rho_{d, gas}$ is the gas density in the mid-plane, and the dynamical pressure of the gas, appropriate to a gaseous halo component, by 
{{}
  $p_{h,gas}=\rho_{h, gas}\sigma^2 =\pi G\Sigma_{gas}\Sigma_d. $
  } Also, $\rho_d$ is taken to be the total disk density in the mid-plane. I can rewrite the star formation rate in terms of gas pressure
\citep{silk09} as 
{ {} 
\begin{equation}
\dot\Sigma_\ast=\epsilon_{SN}\Sigma_{gas}{\sigma\over R_d}=
\Sigma_{gas} { \epsilon_{SN}\over\sigma_d} \sqrt{\pi G p_{d,gas}\over f_g}\sqrt{h\over R_d}
,
\end{equation}
}
where $f_g$ is the gas fraction and $h$ is the disk scale-height. 
The efficiency of supernova momentum  feedback is
\begin{equation}
\epsilon_{SN}=\sigma_d v_c m_{SN}/E_{SN}=0.02
\sigma_{d,10}v_{c,400}m_{SN,150}E_{51}^{-1}.
\end{equation}
Here $\sigma_d=10\sigma_{d,10}\rm km/s$ denotes the velocity dispersion of molecular clouds, 
$v_c=400v_{c,400 }\ \rm km/s$  is the velocity at which a supernova-driven shell enters the momentum-conserving phase of the expansion, $m_{SN}=150 m_{SN,150}\rm M_\odot$ is the mass in stars formed per supernova, and $E_{SN}=10^{51}E_{51}\rm ergs$ is the initial energy of the supernova explosion.   Recall that $\Sigma_d$ is approximately constant for star-forming disks (Freeman's law), as is $\Sigma_{gas}$ for molecular clouds  (one of Larson's laws), although, for star-forming galaxies,  the global value of  $\Sigma_{gas}$ may be higher  by orders of magnitude in extreme cases.

The star formation efficiency, namely the fraction of gas converted into stars per orbital time or $\dot \Sigma_\ast t_{dyn}/\Sigma_{gas}$, is equal to 
$\epsilon_{SN}.$
 Since
$\epsilon_{SN}\propto \sigma_d,$ 
this relation leads to an enhanced efficiency of star formation in merging  galaxies as well as in starburst galaxies where gas turbulence is enhanced..
This  applies to  extreme starbursts \citep{garcia12},  which most likely have merger or AGN-induced turbulent gas motions.

  {  
In what follows, I will explore the  implications of equation (1) for AGN-induced pressure. Since this equation is central to my discussion, I first give a more rigorous derivation, based on the linearised gravitational  instability of  rotating, external-pressure-confined, vertically stratified  polytropic gas disks. }

 {  The maximum instability growth rate in such disks is found to be enhanced by the external pressure, in a study of self-gravitating disks  that are initially in hydrostatic equilibrium \citep{kim12}. The usual derivation of the Schmidt-Kennicutt law
is 
$$
\dot\Sigma_\ast=\epsilon_{SN}
\Sigma_{gas}  \omega_m,
$$
where $\omega_m$ is the maximum growth rate for the non-axisymmetric gravitational instabilities that control molecular cloud growth and star formation, as derived, cf.  \citep{Elmegreen97, Silk97}, and applied, eg. \citep{kennicutt98,Genzel+10}.  One now can generalise the growth rate to the case of an externally pressure-confined disk: 
$$\omega_{m,p}=t_{dyn}^{-1}\left(1+ {2\over\pi}{p_{ext}\over{G\Sigma_0^2}}\right)^{1/2}\sim t_{dyn}^{-1}(2p_{ext} /p_{d})^{1/2}.
$$
The star formation law now reduces to 
$$
\dot\Sigma_\ast=\epsilon_{SN}\Sigma_{gas}{\sigma\over R_d}=
\Sigma_{gas} { \epsilon_{SN}\over\sigma_d} \sqrt{4\pi G p_{ext}}\sqrt{h\over R_d}
.
$$
Note that in the simulations performed to date, the external pressure increase on the gas-rich disk is typically $\sim 1000$ \citep{gaibler12,wagner12}.
One might expect the linear theory result to set an upper limit on the results of the simulations, as non-linear effects such as cloud destruction by ablation need to be included in any analytic discussion. In fact, the
simulations of triggered star formation \citep{gaibler12} show a factor of 3-4 increase in SFR, but the simulations were stopped when the SFR was still increasing.}

 {  Another result from \cite{kim12} is that the characteristic wavelength of the instabilities is reduced by a factor 
$\sim {2p_{ext}} (\pi G\Sigma_0^2)^{-1},$  or $\sim {2p_{ext}} /p_{d}.$ 
This effect augments the supply of molecular clouds that participate in star formation.
}

\section{AGN Feedback}
Accretion onto the central SMBH results in a powerful wind with a broad opening angle or a narrow jet. The latter,  propagating into an inhomogeneous 
interstellar medium, rapidly develops into a cocoon or bubble  once the bow shock that surrounds and precedes the jet thermalises with the ambient medium.

Building on the beautiful results from adaptive grid numerical simulations that are capable of resolving the Kelvin-Helmholtz instabilities that control the interaction of the AGN outflows with interstellar clouds, I use jet-driven outflows \citep{wagner12}  and jet-triggered star formation \citep{gaibler12} as my canonical case in what follows. 

However
the energetics of either type of outflow (jet-driven or wind -driven: for simulations of the latter case, see \cite{wagner13}), expressed in terms of the AGN luminosity,  are inevitably similar, with  the jet being accompanied by a powerful bow shock, apart from geometrical inefficiency factors that I ignore here.

If indeed interstellar pressure (rather than gas density) is the controlling factor in star formation rate, then we can infer the role of an AGN in driving star formation in the circumnuclear disk by evaluating the contribution of the AGN outflow to the interstellar medium (ISM)  pressure, as developed in  \cite{silk09}.
The ISM pressure due to AGN outflows is
\begin{equation}
p_{AGN}= f_E {L_E\over 4\pi R_d^2 c}=
f_E G{{M_{BH}\Sigma_d} \over {M_d \kappa}}
\end{equation}
where the
Eddington luminosity is 
\begin{equation}
L_E={{4\pi GcM_{BH}}\over{ \kappa}},
\end{equation}
the Eddington accretion rate (or the black hole growth rate at Eddington luminosity) is 
\begin{equation}
\dot M_E=\dot M_{BH}/f_E= {{L_E}\over { \eta c^2}}=4\pi {GM_{BH}\over {c\eta \kappa}}
\end{equation}
and the 
 wind outflow rate from the AGN disk  is \citep{kingp03}
\begin{equation}
\dot M_w= f_E {{L_E}\over v_w c}.
\end{equation}
Here $f_E=f_x L_{AGN}/L_E$  and $\kappa$ is the opacity, quantified below and  generally taken to be dominated by electron scattering at small radii (or by dust at large radii) in  the subsequent
applications.
  {  I have introduced an (unknown) AGN efficiency factor $f_x$ that might be less than unity to incorporate AGN-driven shell dissipation.
This would affect the scalings derived above. For example,  a case can be made for energy-conserving winds \citep{faucheregiguere12}.}

  {   The AGN pressure  introduced here drives a wind, in conjunction with any SNe resulting from triggered star formation.  The radial 
dependence of the over-pressured outflow  is described below in what I call the far regime via a solution for  the pressure-driven shell or ring. In so far as the disk is over-pressured,
this  refers to a continuous input of energy via back flow of  the circumdisk  gas  heated by the frustrated jet or wind. This  quasi-uniform turbulent pressure behind the bow shock is clearly seen in the numerical simulations \citep{gaibler12,wagner12,wagner13}. 
}

{
\subsection{The near and far regimes}
Quasar feedback is commonly divided into two modes: the quasar and the radio modes. This is a great simplification of course, but captures the essential physics. The $M_{BH}, \sigma$ scaling relation, and associated black hole growth,  is controlled by the quasar mode. Quenching of star formation by gas ejection from the galaxy potential well is  regulated by the radio (essentially radio jet) mode. 
These modes correspond to near and far regimes, with radio jets, and the associated bow shocks, linking the two.}
  {  The following discussion of the two modes is meant to be illustrative and plausible but could well be modified if different physical models are adopted, for example with regard to the gas fraction or the dust content.}

{\subsubsection{The near mode}
In the near, AGN-dominated, regime, the radiative momentum and any associated wind drives gas away.
The gas is ionised and hot,  and  the dominant opacity is electron scattering. The geometry in the inner region will be that of a hot disk  but  can be approximated here as quasi-spherical.
I assume that the BH outflow self-regulates the gas reservoir that feeds the disk. 
 Hence
balancing the 
AGN momentum $L_E/c$ with  the acceleration needed to eject a gas shell $\bar f_gGM^2R^{-2},$  where I denote the galaxy-averaged gas fraction by $\bar f_g \approx 0.2,$ I obtain the well-known expression \citep{rees98,fabian99,king03,murray05} for the $M_{BH}-\sigma$ 
relation
\begin{equation}
 \sigma^4={4\pi G^2 M_{BH}/{\kappa \bar f_g} }.  
\end{equation}
This agrees in normalisation and slope with the  observed relation, although understanding the dispersion in the relation requires substantially more physics input that up to  now has largely been lacking.
}

{The gas ejected from the inner region does not acquire enough momentum to be ejected from the galaxy and would  eventually cool and refuel star formation, were it not for the onset of jet-driven 
feedback. I will use jets as my template in what follows since these have hitherto been explored in most detail with high resolution simulations, but, as noted previously,  winds may be equally effective. 
}
{
\subsubsection{The far mode}
Consider the jet-driving  of bubbles.  The jet power is  approximated by the Eddington ratio $f_E=P_{jet}/L_E.$
The jet expands into an inhomogeneous ISM and drives a bow shock. The ram pressure sweeps up a shell of dense gas,  entrains clumps and ablates cold gas. The expanding bubble surface allows a large  ram pressure momentum-boost over radiative momentum by a mechanical advantage factor $f_p\sim 10-300$
\citep{wagner12,wagner13},
the numerical simulations finding an outflow efficiency 
\bea
\dot M_{w}v_w^2/L_{AGN}&=&v_w/c\\
&=&
(0.1-0.4) P_{jet}/L_{AGN}.  
\eea
The momentum acquired by clouds relative to jet momentum due to  ram/thermal pressure is $f_p\approx t_3^{1/2},$ where $t_3 =t/1000 \rm yr.$
In this approximately energy-conserving phase, $v_w^2 \propto  P_{jet}  t R^{-3}, $ and one has $ R\propto t^{3/5},$  so that
\begin{equation}
f_p \propto R^{5/6}\propto v_w^{-5/{4}}.
\end{equation}
As $v_w$ falls from its initial value $\sim (0.3-0.1) c$  to the observed outflow velocities of $\sim 10^3$ km/s, one infers that indeed $f_p\sim10-100.$

The fully 3-D numerical simulations demonstrate the validity of this simple approximation,
and supercede earlier discussions that assume spherical symmetry but still provide some insight. For example, 
an alternative way of viewing the boost is if energy of the initial relativistic outflow is conserved, with a fraction $f_i$ of  the initial wind energy going into shell acceleration, generating bulk motion of the swept-up gas at $v_{shell}. $  In this case, the momentum flux is boosted relative to $L_{AGN}/c$ by a factor $f_i v_w/v_{shell},$
where we might expect $v_w\sim 0.1c$ and $f_i\sim 0.5$ \citep{ faucheregiguere12,zubovasking12}. In fact, the actual physics is likely to be rather more complicated, since the final outflow velocity occurs at the onset of the momentum-conserving phase and is a more complex function of halo mass \citep{silknusser10}.

  {  
First,  I give a back-of-the-envelope estimate  for the AGN-induced pressure exerted on clouds in the disk or spheroid, namely}  
\begin{equation}
p_{AGN}=f_pf_E {L_E\over{4\pi R^2 c}},
\end{equation}
where the mechanical advantage factor $f_p$ includes the effects of wind interactions and $f_E$ is the Eddington ratio $L_{AGN}/L_E.$
In what follows, I write $f_{pE}=f_pf_E, $ and for fiducial values, I will define
$f_{pE}=30f_{pE,30}.$

The wind-driven pressure from the entire galaxy is similar:
\begin{equation}
p_w
={\dot M_{wind}v_w\over {4\pi R^2}}  =f_{pE}{{L_E   \over {4\pi R^2 c}} } .
\end{equation}
}

  {  
Next, I derive the pressure more rigorously. Note first that the relevant pressure is ram pressure. I use the similarity solution for
a wind driven by an active region with uniform energy and mass injection \citep{chevalierclegg}. This  spherically symmetric solution, appropriate to a gas-rich spheroid,  has been generalised to the case of
axial symmetry and a wind-driven gas ring \citep{chevalier}, relevant to the case of a gas-rich disk, and the spherical solution has been recently generalised to include the effects of a dark matter  halo \citep{sharma}. However these complications do not significantly  modify the scalings that I use here.  The similarity solution for the wind  shows a smooth transition from subsonic flow at the centre to supersonic flow at large radii.
The energy injection is due to the AGN jet/wind and the mass injection from cloud destruction and ablation. Additional energy and mass injection will come from triggered supernovae.
}

  {  
The similarity solution is expressed  in terms of dimensionless velocity, density and pressure variables:
$$u_\ast =u{\dot M^{1/2}\over \dot E^{1/2}},\ \rho_\ast=\rho{\dot E^{1/2}\over \dot M^{3/2}}R^2 , \  p_\ast={p\over \dot M^{1/2}\dot E^{1/2}}R^2.$$
The ram pressure is $p+\rho u^2$ for the spherical outflow or $p + \rho (u^2 + {5\over 18}v_c^2)$  for the expanding ring, where $v_c$ is the disk circular velocity. The pressure (thermal and ram) as well as the density indeed decrease as $r^{-2}$ outside the active region of radius $R$: all variables are approximately constant within this region, if the star formation rate is proportional to the swept-up mass. This should  be true, as anticipated here,  in the event that  triggered star formation-induced SNe  add to the AGN energy injection in driving the wind.
}

{
 \subsection {Opacities}
 A key question here is what to assume for opacity in the outer region where galaxy-wide outflows are driven. In the absence of significant dust, the electron scattering opacity  is $\kappa_{es}=\sigma_T/m_p=0.38 \ \rm cm^2/g.$
In fact, while electron scattering opacity is appropriate both in the outer protogalaxy and  close to the SMBH, where any dust would be evaporated, the situation is different in the galactic core where the massive black hole forms. There is associated massive star formation, which in most theoretical scenarios is a required precursor to supermassive black hole formation \citep{volonteri12}.  In such situations, there inevitably is metal and dust production during the assembly phase of the central SMBH. Phenomenology demonstrates that cores of spheroids are metal-rich, usually supersolar, and that quasar broad emission line regions are highly enhanced in heavy elements relative to solar values \citep{simonhamann10}. The latter paper argues that the lack of correlation of metallicity with  host galaxy star formation rate suggests that the enrichment occurred at a prior phase, presumably during the coevolution growth phase of the bulges and that of the SMBH. 
 }

{
 I  adopt  a dust opacity
at $T\simlt 200$K given by  the Rosseland mean  opacity that depends only on temperature,  scaling approximately as 
\begin{equation}
\kappa =\kappa_0 (T/T_0)^{\beta}
\end{equation}
where $\kappa_0T_0^{-2}=2.4\times 10^{-4}\rm cm^2 g^{-1}K^{-2},$
and is valid at   $T \simlt T_0\equiv 100$ K \citep{semenov03}. This is appropriate in the far infrared regime, for $h\nu\simlt kT.$  The dust opacity rises steeply with increasing frequency and 
a typical dust opacity in the UV is $\sim 100$gm/cm$^2$.
}

 In the vicinity of the AGN, the dust grains are heated beyond their sublimation temperature. The dust sublimation radius for the most refractory grains, pure graphite cores
 with a sublimation temperature $T_{sub}$ of 1800 K, is 
\citep{kis12}
\begin{equation}
R_{sub}= 0.5 \left({L_{bol}\over{10^{46}\rm ergs/s}}\right)^{1/2}\left(1800\rm K\over T_{sub}\right)^{2.8} \rm pc.
\end{equation}
 Within this region the opacity is dominated, at sufficiently high temperatures, by electron scattering opacity. This should be the case at the inner edge of the nuclear accretion disk.

 I will therefore adopt the following model for opacities in what follows.
 I assume dust opacity is dominant outside the sublimation  region, and normalise to $\kappa=100\kappa_{100}\rm cm^2/gm $  in order to evaluate the balance between AGN luminosity and outflow with protogalactic accretion rate, and electron scattering opacity within. I use the former to compute the 
 momentum balance at large radii, typically a few kpc, and the latter to evaluate the Eddington luminosity  associated with a specified mass accretion rate onto the black hole, typically at 0.1 pc or less.

\subsection{Gas disk}
To proceed further, we need to evaluate the gas surface density.
Let us assume a disk geometry for the gas, following  the discussion of \cite{thompson05}. 
The disk structure is determined by 
\bea
 \Sigma_{gas} =2\rho_{d,gas} h, 
  p_{d,gas} 
=\rho_{d,gas}\sigma_d^2, \cr
 \Sigma_d={\sigma^2\over \pi G R},\ \ 
 \Omega^2=GM_d/R^3,
\eea
where $h=\sigma_d/\Omega$
is the disk scale height, $\sigma_d$ is the gas velocity dispersion,  and the spheroid velocity dispersion is $\sigma=\sqrt{2}  \Omega R .$

We assume the gas disk, embedded in the spheroid,  is massive enough to be self-gravitating 
and maintains itself in a state of marginal gravitational instability so that the Toomre parameter $Q\sim 1.$  
The inner disk will be stable because of AGN heating. High resolution observations of nearby AGN \citep{sani12} indeed suggest the thick disk observed on 
$\sim 100$ pc scales is stable to star formation ($\sigma_d=20-40 \rm \ km/s$ and $Q> 1$).

The outer disk instability, where cooling is significant, is set by 
\begin{equation}
Q={\kappa_\Omega \sigma_d\over{ \pi G \Sigma_{gas}}}, \ \ 
{\rm with} \ \ 
\Sigma_{gas}
={2\Omega \sigma_d\over \pi G Q}, 
\end{equation}
where $\kappa_\Omega=\sqrt{4\Omega^2 + d\Omega^2/d{\rm ln}r}$ is the epicyclic frequency.

 The disk  stellar and gas surface densities  (identifying $M_\ast$ with $M_d$), are
\begin{equation}
\Sigma_\ast={\sigma^4\over{\pi G^2M_\ast}} \ \  {\rm and} \ \ 
\Sigma_{gas}={{\sigma^3\sigma_d}\over{\sqrt{2}\pi Q G^2M_\ast}}.
\end{equation}
Hence the gas fraction is
\begin{equation}
f_g={\sigma_d\over \sqrt{2}Q\sigma}.
\end{equation}

To avoid a detailed radiative transfer model, we adopt a more global approach.
In particular, to compute emission characteristics, we will need to evaluate the radial temperature profile.
This is beyond the scope of this paper.
For generality, however, we give some key equations that, in particular, highlight the role of the opacity
which is used elsewhere in this study  in connection with the Eddington luminosity.

To infer the emission properties, we need to evaluate the temperature and density in the star-forming region.
If the outer disk radiates predominantly by dust emission, one can compute the dust temperature by setting 
\begin{equation}
2\tau_{\kappa} \sigma_{SB} T_{dust}^4 =\epsilon_{rad}c^2 \dot\Sigma_\ast,
\end{equation}
 where $\epsilon_{rad}\sim 0.001 $ is the ratio of radiation per unit mass of newly formed  stars for a typical IMF.
I require the dust optical depth to be of order unity.

Based on   the dust emissivity ($\propto \nu^\beta$) fit to the mean spectral energy distribution of quasar hosts in in the redshift range
$1.8 < z < 6.4$  \citep{beelen06},  I  adopt  $\beta \approx 1.6,$ and
I find the dust disk temperature in the optically thin regime  from the preceding model to be 
\bea
T_d^{disk}
=T_0\left(
{ \epsilon_{rad}\epsilon_{SN}c^2\over {{2}\kappa_0t_{dyn}\sigma_{SB}T_0^4 }}\right)^{1\over{4+\beta}} \cr
= 33{\rm K}\left(\epsilon_{r,-3}\epsilon_{SN,0.02}\over t_{d,7}\kappa_{2.4}\right)^{0.18},
\eea
in reasonable agreement with the data for disks of normal star-forming galaxies.
Here $\epsilon_{SN} =0.02\epsilon_{SN,0.02}$, $\epsilon_{rad}=0.001\epsilon_{r,-3},$ $t_{dyn}=10^7t_{d,7} {\rm yr} =2.10^7 {\rm yr} \ R_{kpc}^{1/2}\Sigma_{d,100}^{-1/2} ,$ 
$\kappa=2.4\kappa_{2.4}\rm cm^2/gm$ at 100K, $R_d=R_{kpc}\rm kpc,$  and $\Sigma_d=100\Sigma_{d,100}\rm M_\odot/pc^2. $
Additional heating will come from the AGN-triggered star formation as discussed below, so that quasar host galaxies will be warmer.

\subsection{AGN self-regulation}

We now assume AGN self-regulation controls the gas reservoir and hence the size of the star-forming disk.
{
In reality it is the innermost disk that is primarily affected and we assume that in this region the rotation curve approximates that of a rigid body. This is of course appropriate for the spheroidal stellar distribution  that is produced as a consequence of the jet-driven bow shock that drives turbulent compression of gas clouds and star formation. The gas velocity dispersion is increased
as a consequence of the jet interaction.
}

{
The region where   AGN feedback can  be positive is determined by the condition that the AGN-induced pressure exceeds the  dynamical pressure that controls the ambient interstellar medium.
{{}
Numerical simulations demonstrate that enhanced AGN-driven pressure from jet backflow
indeed compresses the disk gas and enhances star formation \citep{gaibler12}. The result is not surprising despite the apparent miss-match  in time-scales between  AGN pressure pulses and star formation, because it is the local star formation time-scale in over-pressurised dense gas clumps that is relevant rather than the global disk time-scale.
}
The dynamical pressure is 
\begin{equation}
p_{dyn}=\pi Gf_g \Sigma_d^2,
\end{equation}
and the AGN-induced  pressure is
\begin{equation}
p_{AGN}=f_{pE}{L_E\over{4\pi r^2c}}=f_{pE}{GM_{BH}\over {\kappa r^2}}.
\end{equation}
{
Note that here $f_g$ is a function of radius, whereas in the expression that we use for the black hole mass, 
{{}\begin{equation}
M_{BH}=\bar f_g{{\kappa\sigma^4}\over{4\pi G^2}}={\kappa\over 4}\bar f_g \pi R_d^2\Sigma_d^2,
\end{equation}
}the gas fraction is computed from near zone momentum balance and is the mean value for the galaxy.
Here I  have  used  the disk relation $\Sigma_d=\sigma^2/(\pi G R_d),$ and find that 
 the pressure ratio reduces to 
\begin{equation}
{p_{AGN}\over p_{dyn}}=f_{pE}{M_{BH}\over{\kappa\pi r^2 f_g\Sigma_d^2}}
={{f_{pE}\bar f_g}\over{4 f_g}} \left({R_d\over r}\right)^2
\end{equation}
in the gas-rich inner disk.
Moreover the dynamical pressure as estimated above most likely overestimates the true interstellar medium pressure.  I conclude that AGN-induced pressure is likely to play an important role in regulating star formation in the inner disk on kpc scales. At smaller radii,  AGN pressure dominates.
}

{
\subsection{Bulge formation}
I  evaluate the radius within which AGN-induced pressure dominates. I refer to this scale as the (proto)bulge radius, $R_b,$ which I identify  in order of magnitude  with the disk scale length $R_d,$ my assumption being that the newly formed stars within $R_d$ form the bulge. In subsequent numerical examples, I take $R_b\approx 0.1 R_d,$  as an estimate.
The resulting  stellar bulge could form by satellite infall, disk secular instability, or as
argued below, by AGN-induced outflows and star formation.
}

Note that high redshift  compact quiescent galaxies  may indeed contain  disks \citep{chevance12}.
The effective radius  of star-forming galaxies   observed at high redshift 
\citep{mosleh12}
is comparable  to that of classic bulges,  as well as to the scales inferred  for the predecessors  of compact quiescent galaxies at lower redshift \citep{barro12}. 
}

\subsection{Positive feedback: AGN-driven star formation}

My basis hypothesis is that AGN can have positive as well as negative feedback on star formation rates.  The observational consensus is mixed. Certainly,  radio jet-induced triggering does occur, both of star formation rates \citep{croft06} and of molecular gas formation \citep{feain07}. Quenching of star formation is  established for nearby active galaxies \citep{schawinski07}. 

However recent surveys find little evidence that x-ray luminous AGN quench star formation  \citep{harrison12} and indeed that optically selected radio-loud QSOs  have enhanced star formation at lower luminosities \citep{kalfountzou12}. 
{{} The latter result raises the question of why such an effect is not seen at high radio power. One could speculate as follows. The relevance of high Eddington luminosity to positive feedback is observationally elusive. It might be that at low Eddington luminosities, mechanical feedback is dominant, in which case this would plausibly be the major source of positive feedback. A testable prediction is that evidence for enhanced outflows should be especially prominent in positive feedback candidates, targeted by elevated SSFR at high redshift.

In what follows, I will assume that enhanced pressure associated with the central AGN enhances star formation. This effect may be more prominent at lower luminosities as the starburst is likely to saturate at high luminosity, in part due to induced strong outflows. These are observed to be initiated at high pressures and associated mass loading \citep{newman12}.

The AGN pressure-driven star formation rate is
{ {} 
\begin{equation}
\dot \Sigma_\ast^{AGN}=
{\epsilon_{SN}\over\sigma_d} \Sigma_{gas}\sqrt{\pi Gp_{AGN}\over f_g}=
{\epsilon_{SN}\over\sigma_d}\Sigma_{gas} G\sqrt{\pi f_{pE}M_{BH}\over {\kappa}f_gr^2}.
\end{equation}
}
This 
yields a Schmidt-Kennicutt-like  law at a given ratio of black hole to spheroid mass:
{{}
\begin{equation}
\dot \Sigma_\ast^{AGN}
=G\Sigma_{gas}^{3/2}{\pi\epsilon_{SN}\over{f_g\sigma_d}}
 \left(M_{BH}\over M_\ast\right)^{1/2}
 \left({ f_{pE}}\over{\kappa}\right)^{1/2}.
\end{equation}
}

If $\tau=\kappa\Sigma_{gas}/2$ is specified, one can rewrite this as
{{}
\begin{equation}
\dot \Sigma_\ast^{AGN}
={\Sigma_\ast\over\tau_S}\zeta(\tau/2)^{1/2}f_{pE}^{1/2}\left(M_{BH}\over M_\ast\right)^{1/2},
\end{equation}
}
where
{{}$$\zeta=
{{\epsilon_{SN}c}\over {2\sigma_d}}
=
{{m_{SN}v_c c}\over {2 E_{SN}}}$$ 
}
and $$\tau_S = c\kappa(4\pi G)^{-1}.$$

 The corresponding specific star formation rate is
{{}
\begin{equation} 
SSFR= {{\zeta }\over { \tau_{S}} }  \sqrt{ { {\tau f_{pE}\over 2}  } {M_{BH}\over{M_\ast}}}.
\end{equation}
}
 I set  $$\dot M_{BH}=4\pi {GM_{BH}\over{c\eta\kappa}}=M_{BH}/t_{BH},$$
where the black hole e-folding growth time
$$t_{BH}= \eta \tau_S \equiv c\eta\kappa(4\pi G)^{-1}=4.3\times 10^7\eta_{0.1}\rm \ yr$$
 is equal to  the so-called Salpeter time and I define $\tau_S = c\kappa(4\pi G)^{-1}\equiv 4.3\times 10^8\rm yr$ as a reference time (the Salpeter time at 
100\% efficiency). The numerical values assume electron scattering opacity and $\eta\equiv 0.1 \eta_{0.1}.$
I  note that  
$${{m_{SN}v_cc}   \over E_{SN} }
=360{{m_{SN,150}v_{c,400}}    \over   E_{51}}$$ and 
\begin{equation}\zeta=180 {{m_{SN,150}v_{c,400}}    \over   E_{51}}.
\end{equation}

To infer the disk-averaged specific star formation rate (SSFR), let $ \tau =\bar\tau={f_{pE}\over 2}{M_{BH}\over M_\ast}$ at $r=R_d.$
 The SSFR is given by 
{{}
\bea
SSFR&=& {{\zeta }\over { \tau_S} }  {M_{BH}\over{M_\ast}}{f_{pE}\over 2}\\
&\sim& 
{3\over \tau_S}
\zeta_{200}{M_{BH}\over{10^{-3}M_\ast}}f_{pE,30}
\\ &\sim &
(10^8 \rm yr)^{-1}, 
\eea
}
since $M_{BH}\sim 10^{-3} M_\ast,$  
where $\zeta=200\zeta_{200},$ and $f_{pE}=30f_{pE,30}.$ 
This is similar to what is observed at $z\simgt 2$
\citep{weinmann11, bouwens12, stark13}.
The ratio of stellar luminosity to mass  is 
\begin{equation}
{L_\ast^{AGN}\over M_\ast}=\epsilon_{rad} c^2 (SSFR)\approx {{2.10^{17}\rm s}\over t_{BH}}{\rm erg \  {g}^{-1}s^{-1}} \approx 200{\rm M_\odot\over L_\odot},
\end{equation}
and agrees with ULIRG observations
\citep{scoville03}, although possibly requiring a slightly top-heavy IMF
(with $\epsilon_{rad}\sim 2.10^{-3}$).

We now have the following equation for the AGN-induced star formation rate in terms of the black hole growth rate:
{}
\begin{equation}
 \dot M_\ast ^{AGN} = { \zeta}  {M_{BH}\over \tau_{S}}\sqrt {{\tau f_{pE}\over 2}
  {M_{\ast}\over M_{BH}}}
={\eta \zeta\over 2} f_{pE} \dot M_{BH}  {R_d\over r} .
\end{equation}

I note  that 
{{}\begin{equation}
{ \dot M_\ast ^{AGN}\over \dot M_{BH}} 
= {{\eta \zeta f_{pE}}\over 2}
{{R_d\over r}}.
\end{equation} 
}
This ratio $(\sim 600$) is similar to what is observed for stacked AGN at $z\sim 2,$
if I set   $r\sim 0.5 R_d$    and $f_{pE}\sim 30 $ \citep{mullaney12}.

 The dust temperature is dominated by the inner disk,
 \bea
T_d^{AGN} =T_d^{disk}\left(p_{AGN}\over p_{dyn}\right)^{1\over{2(4+\beta})}= T_d^{disk}\left(R_d\over r\right)^{3\over{4(4+\beta)}}.
\eea
The enhanced dust temperatures $(\sim 45\rm K)$ correspond to those observed for AGN, evaluated at $r\sim 0.1R_d$ or $\sim 100$ pc.

\subsection{Gas accretion and star formation rate}
%

%

The ratio  of black hole growth  to stellar mass growth time-scales, can be written
{{}
\bea
SSFR . t_{BH}={{\eta\zeta f_{pE}}\over 2}
{M_{BH}\over M_\ast}\left(R_d\over 2r\right)  \cr
\approx 0.3\eta_{0.1}\zeta_{200}
\left({10^{3}M_{BH}}\over M_\ast \right).
\eea
}
The stellar mass grows over a similar time-scale to that of the black hole: our model couples stellar mass growth and black hole growth. 
Note however that the model does not provide a duty cycle:  it does not distinguish between a single period of sustained growth or of  many shorter periods.

Evidence for an exponentially rising burst of star formation may arise in the form of the flattening or bluening of rest frame UV continuum slopes of star-forming galaxies at high redshift ($z\sim 4$ for LBGs  \citep{jones12}).  Another possible indicator comes from the frequency of starbursting galaxies at $z\sim 2$: these only represent a significant fraction 
(of order 50\%) of the SFR main sequence if the starburst time-scale is as short as 
$\sim 20$ Myr rather than the customarily adopted $\sim100$ Myr \citep{rodighieri12}. 

Dwarf galaxies undergo short, intense starbursts on similar short time-scales \citep{weisz12}. If these episodes were  AGN-induced, it has been suggested that outflows from intermediate mass black holes may be responsible for ejecting a substantial fraction of baryons from the dwarfs \citep{silknusser10}, something that supernova feedback seemingly fails to accomplish \citep{powell11, peirani12}.

I now make use of
$f_g=\bar f_g (R_d/r)^{1/2}$ 
and integrate the star formation rate induced by AGN, expressed as
{{}
\begin{equation}
\dot M_\ast ^{AGN}  =
{\eta \zeta\over 2} f_{pE} \dot M_{BH}  {R_d\over r} 
={\eta \zeta\over 2} f_{pE}\dot M_{BH}
( f_g/\bar f_{g})^2
\end{equation}
}
for   the redshift dependence of three different gas supply  models. 
The models are:
\begin{enumerate}
\item 
 constant gas fraction as observed at high redshift in star-forming galaxies \citep{bothwell12}, discussed above, so that $\rho_g(r) \propto\rho_\ast(r),$
\item  closed box ($M_g+M_\ast=constant$) as might be appropriate for a major 
merger-induced gas supply,  
\item  cosmological accretion ($M_g=M_{g,0}(1+z)^\gamma$)
 along cold filaments as favoured by cosmological simulations. 
Here $\gamma \approx 2.2$ and $\dot M_g \propto t^{-\alpha}$ with $\alpha\approx 5/3$ for LCDM \citep {neistein08}. I neglect gas sinks due to star formation and outflows, as justified 
by \cite{bouche10}. 
\end{enumerate}
We infer  for all of these models that that   the  stellar mass undergoes exponentially rapid growth on a time-scale of order the black hole growth  time-scale, since the exponential term dominates all solutions for $M_\ast.$ 

\subsubsection{Constant gas fraction}
At constant $f_g$ (case 1),  the stellar mass integrates to 
 \bea
  M_\ast ^{AGN}  
 ={\eta \zeta} f_{pE}(f_g/ \bar f_g)^2 M_{BH}  \cr
 =3000\eta_{0.1}\zeta_{200} f_{pE,30}(f_{g,0.5}/ \bar f_{g,0.2}) ^{2}M_{BH}. 
 \eea
The ratio of star formation to black hole growth rates is approximately equal to the ratio of stellar mass to BH mass produced during the coeval growth phases.
Both the stellar mass in the starburst and the star formation rate increase exponentially, on the black hole growth time-scale.

\subsubsection{Closed box}
In the closed box model, case (2),    I define $ M_0=M_g + M_\ast$,  $\mu =M_\ast/M_0$ and $f_g=M_g/M_\ast.$
The star formation rate is 
{{}
\bea
 \dot M_{\ast}^{AGN} 
&=&
{\eta\zeta\over 2}\dot M_{BH}{f_{pE}(f_g/\bar f_g})^{2} \\
& =&
 {\eta \zeta}{\dot M_{BH}}{f_{pE}/\bar f_g}^2 (\mu^{-1}-1)^2.\eea
 }

The solution is
{{}
\begin{equation}
{M_{BH}\over M_\ast}=
{2\over{\eta\zeta f_{pE}}}{1\over\mu}\left[\mu-1 +{1\over{1-\mu}}+ 2\ln{(1-\mu)}\right].
\end{equation}
}
In the early, gas-rich limit, this reduces to 
{{}
\begin{equation}
{M_{BH}\over M_{gas}}={2\over{\eta\zeta f_{pE}}}{f_g^2\over 3}.
\end{equation}
}
In the late, gas-poor limit, we have 
{{}
\begin{equation}
{M_{BH}\over M_{\ast}}={2\over{\eta\zeta f_{pE}}}f_g
=0.001 \eta_{0.1}^{-1}\zeta_{200}^{-1}f_{pE,30}^{-1}\bar f{g,0.3}.
\end{equation}
}
Here I denote the current epoch gas fraction ($f_{g,0}$) with $\bar f_g.$
\subsubsection{Accretion}
In the accretion model, case (3), the star formation rate is 
{{}
\begin{equation}
\dot M_{\ast}^{AGN} 
 ={{\eta \zeta}\over 2}f_{pE}{\dot M_{BH}}{ f_{g,0}/ \bar f_g}  (t_0/t)^{2\alpha}.
\end{equation}
}
This integrates to ($t_0$ is the Hubble time)
{{}
\begin{equation}
{M_\ast\over M_{BH}}={{\eta \zeta}\over\alpha}{ f_{pE}} \   (t_0/t_{BH})^{2\alpha} g(t),
\end{equation}
}
where $x=t/t_{BH}$ and $g(x)$ is a weakly varying function (of time)
{{}
$$g(x)=e^{-x}\int e^x x^{-2\alpha}dx. $$
}
In all cases, the ratio of black hole to bulge mass is approximately equal to the instantaneous ratio of black hole accretion rate to star formation rate and of order 
$$2 (\eta\zeta f_{pE})^{-1}(\bar f_g/f_g)^{2}
\sim 0.001$$  
for typical parameter choices ($ \eta\sim 0.1, f_{pE}\sim 30, \bar f_g =0.2, f_g\sim 0.4, 
\  m_{SN,150}\sim v_{400}\sim E_{51}\sim 1$).

\subsection{AGN-driven outflows}
Black hole growth is generically expected to generate outflows. The outflow rate is given  by
\bea
 \dot M_w
 =f_{p}{\eta\over v_w c} L_{AGN}
 &=&f_{pE} {c\over v_w}\dot M_{BH}\\
 &=&1000f_{pE, 30}v_{w,10^4}^{-1}\dot M_{BH}
\eea
since $L_{AGN}=f_E\eta c^2 \dot M_{E}.$  Here the initial wind velocity $ v_w=
10^4 v_{w,10^4}\rm km/s.$ Incorporating the approximate dependence of the momentum advantage factor $f_p$ on $v_w$  as discussed above, 
 one infers that 
$\dot M_w \simpropto L_{AGN}\  v_w^{-11/6}.$ Simulations show that observed outflow velocities match this prediction \citep{wagner12}.

One can now express the outflow rate in terms of the star formation rate:
{{}
\begin{equation}
\dot M_w=
2{c\over v_w}{1\over{\eta\zeta}}
(\bar f_g/f_g)^2
\dot M_\ast^{AGN}.
\end{equation}
}
Hence the ratio of outflow to star formation rate is
{{}
\begin{equation}
{ \dot M_w\over \dot M_\ast^{AGN}}=
2{c\over v_w}{1\over{\eta\zeta}}
(\bar f_g/f_g)^2
={{0.75}\over{\eta_{0.1}\zeta_{200}v_{w,10^4}}}
\left({\bar f_{g,0.2}\over{ f_{g,0.4}}}\right)^2.
\end{equation}
}
Even in the most massive galaxies, the outflow rate can be of order the star formation rate, due to the role of the AGN.

Empirically, there is a generic coupling for  AGN between star formation rate and  outflow rate,  e.g. as found for the well-resolved case of Mrk 231
\citep{rupke11}.  This coupling is also true for star-forming galaxies without strong AGN \citep {bouche12}. However the difference in the latter case is that the outflow velocities are  of order the circular velocities and well below the escape velocities. This result suggests  that supernova-driven outflows  drive gas circulation in the halo but cannot eject significant  amounts of baryons from the galaxy potential wells. On the other hand, the outflow velocities associated with luminous AGN generally exceed the escape velocity of the host galaxy.

\section{Summary and discussion}
  {   Triggered star formation is not a new concept. It is well-studied in the galactic context for triggering by massive star HII regions, first introduced by \cite{elmegreenlada}, and since widely explored on galactic scales. Triggering of massive star cluster formation commonly occurs  in galaxy mergers \citep{whitmore10}.
 For AGN, however, the idea is somewhat new, and has attracted increasing interest in view of the complex  and hitherto obscure interaction between AGN and star formation. }

  {  
Major mergers are often blamed for both enhanced AGN activity and star formation.  
{There is an active ongoing debate however:  for example, it has been argued that  
new-born spheroids are not the product of major mergers \citep{kaviraj13}, nor is enhanced star formation in massive galaxies at $z\sim 0.6$  primarily due to interactions \citep{robaina09}.  There is   a general consensus  however that while major mergers are nevertheless important for the most extreme star formation rates observed at high redshift, and are important locally in accounting for ULIRGs, they are likely subdominant at high redshift both  in triggering the bulk of star formation \citep{kartaltepe12} and in triggering  AGN \citep{treister12}.}


One alternative to mergers  is gas accretion triggering. But for formation of early-type galaxies,  accretion is generically filamentary according to numerical simulations, e.g. \cite{sales12}.  However observations reveal quasispherical star formation-induced gas excitation around radio quiet AGN \citep{liu13}.  This is  suggestive of  AGN triggering by quasispherical winds from AGN rather than by accretion flows. 

While  I have focussed here on radio jet-driven bubbles, being guided by the results from recent numerical simulations, much of this discussion is applicable to the radio-quiet mode, where quasar-driven winds provide similar feedback. Indeed a recent numerical study confirms the similarity between jet and wind-induced outflows \citep{wagner13}, with corresponding implications for pressure enhancements and (presumably) triggered star formation according to the scheme developed in \cite{gaibler12}.  A possible example of a QSO-wind induced triggering of a young galaxy at $z=3.045$ is given by \cite{rauch13}.

  {  
In summary,  I have argued that AGN triggering of star formation  arises via AGN pressure-regulation. This  allows  inclusion of AGN-induced pressure into what is essentially a reinterpretation of the usual star formation law, by introducing jets and/or winds into an inhomogeneous interstellar medium,  and leads to enhanced star formation rates. }

Three possibilities are considered for the evolution of the gas fraction: a constant gas fraction (for illustrative purposes),  and the more realistic cases of a closed box model (applicable in the case of a major merger), and an accretion merger (relevant for minor mergers or filamentary accretion of cold gas.  The latter case is favoured if indeed a significant fraction of SMBH growth occurs predominantly in disks \citep{schawinski12}.

Simple expressions are derived that relate black hole growth, star formation and outflow rates. 
The specific star formation rate is found to be essentially identical to the specific black hole accretion rate. Since the latter must be on the order of $3.10^{-8}\rm
yr ^{-1}$ in order to grow supermassive black holes by $z\sim 7$ (as reviewed recently by \cite{volonteri12}),
this means that starburst time-scales are typically a few tens of millions of years rather than the normally adopted $\sim 10^8$ yr. 

One consequence is that starbursts are a major contributor to star formation and are predicted to  lie above the galaxy main sequence, as observed by Herschel at $z=1.5-2.5$ \citep{rodighieri12}. The black hole mass is found to be around 0.001 of the old (spheroid) stellar mass. The black hole accretion rate is a similar fraction of the star formation rate,  and predicted to be a factor of 2 or so higher, i.e. $\sim 0.002, $ in the filamentary gas  accretion model, as inferred when stacking x-ray selected samples of ultraluminous AGN at high redshift \citep{mullaney12}.

The ratios of black hole to spheroid mass and  of the comoving  black hole accretion  rate density to star formation rate density are found to track each other as a function of time, although offset by $\sim 1000.$ This is well known at low redshift \citep{silverman09} but the present model predicts that a similar offset continues at high redshift.

 Indeed, allowance for the prevalence of buried luminous AGN  \citep{imanishi10, treister11}
 flattens the observed black hole accretion rate density at high redshift. Allowance for lack of dust at high redshift  ($z\sim 4-7$ from  \cite{bouwens12}), and especially top-down galaxy  formation \citep{wechsler12},  lowers the star formation rate density at high redshift. I conclude that the shapes of the two  rates may well be  similar over $z=0-6$, but offset by a factor of order 1000.

Both black hole and star formation rates are exponentially increasing functions of time. This lowers the mean age of luminous starbursts and should lead to flatter UV continuua than in any unaccelerated burst, as would be the case in the absence of positive AGN feedback. Systematically flatter rest-frame  UV are indeed observed at $z\sim 4$ \citep{jones12}.

Exponentially increasing star formation rates are  also found in certain wind-regulated hydrodynamic models and are a consequence of high cold gas accretion rates at early epochs \citep{finlator11}. These models give 
a wide range of fits to the SSFR \citep{dave11, schaerrer12}. However the rate of early  cold accretion in massive galaxies may be severely overestimated according to more realistic, moving mesh hydrodynamic simulations \citep{nelson13}. The models advocated here provide an alternative means of obtaining rapidly rising star formation rates.


 The exponential star formation rate self-limits the period of black hole growth, since the gas reservoir will be depleted. The limiting star  density in the starburst is found empirically to be of order \citep{hopkins10}
$10^{11}\rm M_\odot/kpc^2.
$
To  attain such a high value, one might need black hole growth, as well as spheroid growth, to occur via a series of short bursts with a duty cycle of order 10\% over a period of perhaps a Gyr. These latter parameters, not calculated here but the subject of future work, are likely related to the radiative efficiency of mass accretion onto the black hole.

With regard to massive galaxies, the problem with the inability of  a single mode of star formation to reproduce the mass function of galaxies at both low and high redshift is well known. Models that fit the high redshift mass function fail at low $z$ \citep {henriques11}, and models that fit at low $z$ fail at high $z$ \citep{fontanot09}, especially for the most massive galaxies \citep{mutch12}. 

There is a strong case for introducing a second mode of star formation that provides higher efficiency, in particular at high redshift. Whether this is due to tweaking the conventional density threshold and Schmidt-Kennicutt law approach,  for example by appealing to enhanced gas accretion as might be supplied in mergers \citep{kochfar11} or by introducing new physics associated with positive feedback from AGN,  as advocated here, has yet to be determined.

  
    {  
 Finally, I  make several speculative suggestions for possible observational tests of positive feedback by AGN. }
 
\begin{enumerate}
\item 
  {  
It is clear that high resolution observations of massive young galaxies and of molecular gas at high redshift will help elucidate these issues.
One target would be the enhanced star formation and turbulence  induced at the edges of cloud complexes. This is seen on cloud scales  in nearby star-forming clouds \citep{dirienzo12}. With ALMA or GMT/TMT/ELT resolution, it would be interesting to extend such studies to 
 the massive star forming clumps seen in redshift $z\sim 2$ galaxies, which include extremely  high star formation densities   
\citep{blain13}.
}

\item
   {  One specific example is the quasar host galaxy \citep{walter09}, and another is a galaxy at $z=6.34$ with the most extreme star formation rate surface density yet recorded  \citep{riechers13}.
These so-called hyper-starbursts at high redshift provide sites where evidence of an AGN trigger might be sought.  }

 \item 
A class  of objects where positive feedback is inferred and further evidence could be sought 
for evidence of mechanical driving is that of  AGN in the early universe  with enhanced far infrared emission lines ([CII], [OIII])  from PDRs on the surfaces of molecular clouds. These observations are indicative of localised (kpc-scale) intense star formation bursts \citep{stacey10}.

\item 
   { Formation of dense molecular  clouds is a crucial step in the triggering pathway. One manifestation would be strong shocks and enhanced molecular cooling in molecules such as $H_2O$, which is found to be exceptionally strong in high redshift star-forming galaxies \citep{omont13}.
With larger samples, a correlation could be sought between $H_2O$ emission and AGN activity. I note that a lack of correlation is reported for PAH emission \citep{rawlings13}, but the  shorter lived phase of intense molecular cooling is a more robust diagnostic of cloud formation and compression, the necessary prerequisites for star formation.
}

\item 
    {  The enhancement of star formation in the presence of AGN is accepted, as is the occurrence of strong outflows. However any causal link is 
 yet to be established. Stellar populations are found to be older in AGN hosts at $z< 1$ \citep{vitale13}.  This extends previous work on similar 
 age offsets at $z\simlt 0.1$ \citep {schawinski07}. There is evidence for bimodal stellar populations in luminous radio galaxies at high redshift, suggesting that a massive starburst associated with the onset of radioactivity may have occurred \citep{rocca13}.
  It would be interesting to see if stellar population age in AGN hosts were correlated with stellar mass and hence SMBH mass, as triggering might suggest. This is because  the strength of  feedback, as characterised by the Eddington ratio, increases with increasing SMBH mass. 
  }
  
 \item   
     {  It now appears that the comoving AGN accretion rate density tracks the cosmic star formation history, although reduced in mass flux by a factor $\sim 1000,$ once allowance is made for galaxy luminosity downsizing and the frequency of buried AGN. This result needs to be verified with larger samples of AGN and statistical differences sought between the low redshift sample ($z\simlt 2$) where star formation declines with time, and the high redshift sample ($z\simgt 2$), where star formation increases with time. The present model leads one to expect that quenching dominates at low $z$ and triggering at high $z.$   This is a natural consequence of the present model: early positive feedback is followed by a phase of negative feedback. This is difficult to prove for individual objects, especially if the positive feedback phase is both brief and buried. However large samples
observed in these respective redshift ranges  might be expected to reveal  systematic differences in star formation history and gas depletion timescales  of the host galaxies. 
   }

    {  
   \item 
   The scatter in the correlation between black hole mass and spheroid velocity dispersion or spheroid mass  could contain the imprint of AGN triggering. For example, one might expect 
 the significance of   triggering to correlate with Eddington ratio, in which case the residuals in stellar mass would correlate with Eddington ratio and with black hole growth. 
   }
   
   \item
    {  Metallicities  also provide a potential probe. The strong outflows associated with positive feedback will lead to early enrichment of the circumgalactic medium, with enhanced $[\alpha/Fe]$ as a possible chronometric signature. A time-delay between AGN-triggering and enrichment would be expected. Ejection from the galaxies of metal-enriched hypervelocity stars would survive as possible witness to these early events, cf. \cite{silk12}. 
  }   
  
  \item
  { 
  Jet-induced triggering should result in rings of young stars, embedded in rings of compressed molecular gas.  ALMA resolution 
  should probe the morphology of the molecular gas. These rings may fragment to form star clusters: intriguingly there is a possible example of a super-star cluster  with a high velocity relative to the halo gas in a ULIRG \citep{holt09}. More realistic geometries might include chains of sequential star formation triggering.
  }
\end{enumerate}
\section*{Acknowledgements}

I thank Y. Dubois, V. Gaibler, M. Krause, M. Lehnert, C. Norman, A. Nusser  and M. Volonteri for pertinent discussions.  This research has been supported at IAP by  the ERC project  267117 (DARK) hosted by Universit\'e Pierre et Marie Curie - Paris 6   and at JHU by NSF grant
OIA-1124403.

\end{document}